\documentclass[12pt,preprint]{aastex}

\shorttitle{Recent Star Formation in the Leading Arm of the Magellanic Stream}

\begin{document}

\title{Recent Star Formation in the Leading Arm of the Magellanic Stream}

\author{Dana I. Casetti-Dinescu\altaffilmark{1,2}, 
Christian Moni Bidin\altaffilmark{3}, 
Terrence M. Girard\altaffilmark{2}, 
R\'{e}ne A. M\'{e}ndez\altaffilmark{4}
Katherine Vieira\altaffilmark{5}, 
Vladimir I. Korchagin\altaffilmark{6} and 
William F. van Altena\altaffilmark{2}}

\altaffiltext{1}{Department of Physics, Southern Connecticut State University, 501 Crescent St., New Haven, CT 06515, USA, casettid1@southernct.edu}
\altaffiltext{2}{Astronomy Department, Yale University, 260 Whitney Ave. , New Haven, CT 06511, USA, dana.casetti@yale.edu,terry.girard.@yale.edu,william.vanaltena@yale.edu}
\altaffiltext{3}{Instituto de Astronom\'{i}a, Universidad Cat\'{o}lica del Norte, Av. Angamos 0610, Antofagasta, Chile, chr.moni.bidin@gmail.com}
\altaffiltext{4}{Departmento de Astronom\'{i}a,  Universidad de Chile, Casilla 36-D, Santiago, Chile, ramendez.uchile@gmail.com}
\altaffiltext{5}{Centro de Investigaciones de Astronom\'{i}a, Apartado Postal 264, M\'{e}rida 5101-A, Venezuela, kvieira@cida.ve}
\altaffiltext{6}{Institute of Physics, Southern Federal University, Stachki st. 124, 344090, Rostov-on-Don, Russia, vkorchagin@sfedu.ru}

\begin{abstract}
Strongly interacting galaxies undergo a short-lived
but dramatic phase of evolution characterized by enhanced star formation, 
tidal tails, bridges and other morphological peculiarities. The nearest example of a pair of 
interacting galaxies is the Magellanic Clouds, whose dynamical 
interaction produced the gaseous features known as the  Magellanic Stream trailing 
the pair's orbit about the Galaxy, the Bridge between the Clouds, and the Leading Arm, 
a wide and irregular feature leading the orbit. Young, newly formed stars in the Bridge 
are known to exist, giving witness to the recent interaction between the Clouds.
However, the interaction of the Clouds with the Milky Way is less well understood.
In particular, the Leading Arm must have a tidal origin, however no
purely gravitational model is able to reproduce its morphology and kinematics.
A hydrodynamical interaction with the gaseous hot halo and disk 
of the Galaxy is plausible as suggested by some models and supporting neutral hydrogen 
(H~I) observations.
Here we show for the first time that young, recently formed stars exist in the Leading Arm,
indicating that the interaction between the Clouds and our Galaxy
is strong enough to trigger star formation in certain regions 
of the Leading Arm --- regions in the outskirts of the Milky Way disk ($R\sim 18$ kpc), far away from the Clouds
and the Bridge. 
\end{abstract}


\section{Introduction}

The Magellanic Clouds (MCs) offer a unique opportunity to study galaxy interactions in unprecedented detail 
due to their proximity to the Milky Way (MW). Thus, detailed mapping of their gaseous content, the 3D kinematics of 
their stellar content, and  the chemical-abundance makeup of these components are readily available for the Clouds. 
The most obvious features of their interaction are the H~I 
structures known as the $\sim 200^\circ$-long Magellanic
Stream (MS), the Bridge, and the Leading Arm (LA) (Nidever et al. 2010). The recent work on the
modeling of the Clouds' interaction by Diaz \& Bekki (2012) makes a compelling case for tidal model,
where the MS, Bridge and LA are made primarily of material pulled out from the SMC during 
two close encounters between the two Clouds. The first encounter took place 
$\sim 2$ Gyr ago, and the second $\sim 200$ Myr ago. This work used
the most recent absolute proper-motion determinations for the Clouds: one HST-based (Kallivayalil et al. 2006), the other
ground-based (Vieira et al. 2010). Both determinations imply exactly two encounters between the Clouds to 
reproduce the MS, LA and Bridge.
As for their motion relative to the MW, HST measurements(Kallivayalil 2006, 2013) favor the scenario where 
the MCs are on the first passage about the MW, with the
MS and LA determined solely by the tidal interaction between the Clouds. 
The ground-based proper-motion measurement allows
for two pericentric passages of the Clouds about the MW in the past 2.5 Gyr, 
and thus some tidal influence of the Galaxy in the formation of the MS, LA and Bridge is expected. 
The main drawback of the tidal models is that, while they produce a leading arm, 
all fail to reproduce the observed multi-branches morphology of the LA, and its kinematics.
A model by Diaz \& Bekki (2011) that also includes a hydrodynamical
interaction of the LA with the diffuse, hot gaseous halo of the MW, better reproduces the 
kinematics along the LA, but not its morphology.
The LA has a complex structure, possibly made of as many as four substructures according to For et al. (2013) and Venzmer et al. (2012), 
situated
above and below the Galactic plane, and encompassing $\sim 60^\circ$ in width. 
It has been argued that there is a strong drag exerted by the MW gaseous 
disk on the H~I substructures in the LA (McClure et al. 2008, Venzmer et al. 2012). 
This is implied by the head-tail velocity
structure of the H~I clouds in the LA, as well as the velocity gradient seen 
in a given LA substructure/arm (Venzmer et al. 2012).
Thus it would be enlightening to search for newly formed stars
in the LA, an expected result of the hydrodynamical interaction between the MS gas and 
the MW gaseous disk and halo. We also note that the most recent ($\sim 200 $ Myr ago) encounter between the two Clouds
which created the Bridge, is abundantly accompanied by recent star formation, a fact
well known since the work by Irwin et al. (1990) and subsequent follow-up by, e.g., Demers et al. (1998).

In a recent study Casetti-Dinescu et al. (2012) listed 567 OB-type star candidates in 
a $\sim$ 7900 square degree area
encompassing the periphery of the Clouds, the Bridge, the LA, and part of the MS.
The photometric and proper-motion selection was aimed at finding hot (earlier than B5) and
distant stars. Also, the proper-motion selection was aimed at selecting stars with 
motions consistent with membership to the Magellanic system.
In the LA region, three stellar overdensities were found comprising a 
total of 45 candidates.
This is a lower limit of such candidates, since the study is area-wise incomplete (Casetti-Dinescu et al. 2012).
Here, we have spectroscopically observed 42 of the 45 candidates. Their  
spatial distribution is shown in Figure 1.
Also shown is the H~I distribution
from the GASS survey
(McClure-Griffith et al. 2009, Kalberla et al. 2010)
for which we have restricted the velocity with 
respect to the Local Standard of Rest,
to be $ 150 \le V_{LSR} \le 400$ km/s. The three candidate overdensities we label:
A at $(\Lambda_M, B_M) \sim (15^\circ,-22^\circ)$,
B at $(\Lambda_M,B_M) \sim (42^\circ,-8^\circ)$, and C at $(\Lambda_M,B_M) \sim (52^\circ,28^\circ)$.
In what follows, we dscribe the spectroscopic observations and the results.

\section{Observations}

Intermediate-resolution spectra were obtained with the IMACS spectrograph
on the 6.5m Baade telescope at Las Campanas Observatory. The setup gave a 
resolution of 1.3$~\AA$ (R$\approx$3500) in the range 3650 to 5230~$\AA$.
The 1200 l/mm grating at the f/4 camera was 
employed at first order, with a blaze angle of $17^\circ$ and 0.75''-wide slit, for a resulting 
resolution of 1.3$~\AA$ (R$\approx$3500) in the range 3650 to 5230~$\AA$.
The average seeing during observations was 0.7'', and the resulting spectral 
The spectral signal-to-noise ratio was higher than 50 for all the targets.
Cross-correlation techniques (Tonry \& Davis 1979)
as implemented in the IRAF {\it fxcor} task, 
were used to measure heliocentric radial velocities (RVs). 
In absence of a prior knowledge of the exact temperature and gravity of the targets, 
the synthetic spectrum (Munari et al. 2005) of a main-sequence B-type star was adopted as the template
\footnote{A mismatch between the parameters of the template and object spectra enhances the uncertainties 
but does not affect the results, especially for hot stars (Moni Bidin et al. 2011).}.
The final uncertainty, taking into account the relevant sources of
errors, are estimated to be between 3 and
14~km~s$^{-1}$, typically $\approx$5~km~s$^{-1}$ for most of the targets.
The spectra are also fitted with standard routines (Bergeron et al. 1992, Saffer et al. 1994, Napiwotzki et al. 1999)
to derive the temperature, gravity, surface helium abundance, and, in some cases, rotational velocity\footnote{Rotational velocities
were not fitted for, but rather used as an input parameter. By using different inputs, the one that gave the lowest chi square, was 
adopted as the value of the $vsini$. These values should be regarded as indicative, with errors $\sim 30$ km~s$^{-1}$}.
At these colors, our major source of contamination is foreground
subdwarf O and B stars (sdBs) and white dwarfs.
Close binaries are extremely common among sdB's (Maxted et al. 2001), hence, RVs alone are not conclusive to assess
the membership of our targets to the LA. To distinguish main sequence stars from
sdBs, besides the surface gravity, we can also use the surface helium abundance because the 
atmosphere of sdB's in the temperature range $T_{eff} > 11500$~K is depleted of helium by 
a factor between 10 and 100 due to
gravitational settling (Baschek 1975, Moni Bidin et al. 2012). Rotational velocity is also 
indicative, as fast
rotators are common among early-type main sequence stars, but not among sdB's (Geier \& Heber 2012).

\section{Results}

Of the 42 candidates observed, we find 19 young, massive stars, together with 22 foreground 
sdB and white dwarf stars, and one uncertain object. 
The density of young stars in regions A and B is higher than in C, at a 
significance of $2.6\sigma$ in A, for instance, after correcting for
areal incompleteness.

We adopt RV = $150$~km~s$^{-1}$ as the lower limit for kinematical membership to the LA
based on H I velocities in the LA (e.g., Venzmer et al. 2012). Note that, in this region of the sky, heliocentric and 
LSR RVs differ by a very small amount (at most 14~km~s$^{-1}$ in region C). 
We find a total of six stars with RV $ > 150$~km~s$^{-1}$: four in region B, two in region A, and 
none in region C (Fig. 1). 
These are listed in Table 1, along with one other star of interest.

\begin{table}[tbh]
\caption{Spectral Parameters for Stars of Interest}
\begin{tabular}{llllllllll}
\hline
 \multicolumn{1}{c}{ID} & \multicolumn{1}{c}{SPM ID} & \multicolumn{1}{c}{$\it l$} & \multicolumn{1}{c}{$\it b$} &
\multicolumn{1}{c}{RV} & \multicolumn{1}{c}{vsini} &
\multicolumn{1}{c}{$T_{eff}$} & \multicolumn{1}{c}{$log~g$} & \multicolumn{1}{c}{$log\frac{N_{He}}{N_H}$} & \multicolumn{1}{c}{Sp. Type} \\
 & \multicolumn{2}{c}{($^o$)} & \multicolumn{2}{c}{(km s$^{-1}$)} & \multicolumn{1}{c}{$(K)$} & \multicolumn{1}{c}{(dex)} & 
\multicolumn{1}{c}{(dex)} & \\
\hline
\multicolumn{1}{l}{B02} & \multicolumn{1}{l}{$2880084481$} & \multicolumn{1}{r}{$275.3$} & \multicolumn{1}{r}{$10.5$} & \multicolumn{1}{r}{$168\pm4$} & 
\multicolumn{1}{r}{$0$} & \multicolumn{1}{r}{$16000\pm400$} & \multicolumn{1}{r}{$3.76\pm0.09$} &
\multicolumn{1}{c}{$[-1]$} & \multicolumn{1}{l}{B4III} \\
\multicolumn{1}{l}{B03} & \multicolumn{1}{l}{$2890097375$} & \multicolumn{1}{r}{$277.7$} & \multicolumn{1}{r}{$13.1$} & \multicolumn{1}{r}{$229\pm5$} & 
\multicolumn{1}{r}{$0$} & \multicolumn{1}{r}{$17500\pm1100$} & \multicolumn{1}{r}{$3.86\pm0.18$} &
\multicolumn{1}{l}{$-1.60\pm0.18$} & \multicolumn{1}{l}{B3IV} \\
\multicolumn{1}{l}{B14} & \multicolumn{1}{l}{$2310104323$} & \multicolumn{1}{r}{$288.8$} & \multicolumn{1}{r}{$12.2$} & \multicolumn{1}{r}{$207\pm5$} & 
\multicolumn{1}{r}{$300$} & \multicolumn{1}{r}{$15800\pm700$} & \multicolumn{1}{r}{$3.60\pm0.15$} &
\multicolumn{1}{c}{$[-1]$} & \multicolumn{1}{l}{B4/5III} \\
\multicolumn{1}{l}{B15} & \multicolumn{1}{l}{$2310425880$} & \multicolumn{1}{r}{$290.8$} & \multicolumn{1}{r}{$9.6$} & \multicolumn{1}{r}{$248\pm5$} & 
\multicolumn{1}{r}{$0$} & \multicolumn{1}{r}{$14800\pm700$} & \multicolumn{1}{r}{$4.17\pm0.15$} &
\multicolumn{1}{l}{$-2.10\pm0.33$} & \multicolumn{1}{l}{sdB} \\
\multicolumn{1}{l}{A15} & \multicolumn{1}{l}{$0400042068$} & \multicolumn{1}{r}{$300.7$} & \multicolumn{1}{r}{$-11.8$} & \multicolumn{1}{r}{$166\pm6$} & 
\multicolumn{1}{r}{$280$} & \multicolumn{1}{r}{$13400\pm600$} & \multicolumn{1}{r}{$3.67\pm0.18$} &
\multicolumn{1}{c}{$[-1]$} & \multicolumn{1}{l}{B7III} \\
\multicolumn{1}{l}{A19} & \multicolumn{1}{l}{$0420365194$} & \multicolumn{1}{r}{$309.4$} & \multicolumn{1}{r}{$-14.1$} & \multicolumn{1}{r}{$234\pm6$} & 
\multicolumn{1}{r}{$260$} & \multicolumn{1}{r}{$17200\pm1900$} & \multicolumn{1}{r}{$4.03\pm0.36$} &
\multicolumn{1}{l}{$-1.58\pm0.42$} & \multicolumn{1}{l}{B3IV} \\
\multicolumn{1}{l}{A08} & \multicolumn{1}{l}{$0390226948$} & \multicolumn{1}{r}{$298.8$} & \multicolumn{1}{r}{$-13.9$} & \multicolumn{1}{r}{$71\pm11$} & 
\multicolumn{1}{r}{$0$} & \multicolumn{1}{r}{$43700\pm1200$} & \multicolumn{1}{r}{$4.47\pm0.12$} &
\multicolumn{1}{l}{$-1.20\pm0.39$} & \multicolumn{1}{l}{O6V} \\
\hline
\multicolumn{9}{l}{The [] indicate an adopted rather than fitted value.}
\end{tabular}
\end{table}

Remarkably, of the six stars
with RV $ > 150$~km~s$^{-1}$, five are young, massive objects, as inferred from $T_{eff}, log~g$ and 
helium abundance.
Three of these stars are also fast rotators. 
The RV average and 
dispersion of these five stars is $201\pm14$~km~s$^{-1}$, and $32$~km~s$^{-1}$ respectively.
Even more remarkably, we find one candidate star to be a very hot, main sequence star with spectral type O6V (Tab. 1), 
and thus a massive ($\sim 40~M_{\odot}$), short-lived (1-2 Myr) star at a heliocentric distance of $\sim 40$ kpc.
In Figure 1, we highlight the 19 young stars, the six stars with RV $ > 150$~km~s$^{-1}$, and the one O6V type star.
We plot the RVs as a function of surface gravity, color-coded by effective temperature, in Figure 2, and highlight
the fast rotators. As a guide, we also indicate the mean RV of the thin+thick disk populations 
and its $\pm2\sigma$ standard-deviation range as derived from the 
Besancon galactic model (Robin et al. 2003).
The group of stars with RV $ > 150$~km~s$^{-1}$ and $log~g \le 4$ stands out in this plot. 

A possible non-LA origin for the young stars must be considered, i.e., that these are runaway stars.
Such stars are believed to 
originate in star-forming regions of the Galactic disk as massive binaries that are disrupted by either a 
supernova, or by a three- or four-body dynamical interaction, 
probably within a young star cluster (e.g., Bromely et al. 2009) 
In principle, all of our 19 young stars could be suspected of having such an origin.
However, the five young B-type stars
with RV $ > 150$~km~s$^{-1}$ do not fit such a scenario, since their RV dispersion of 33~km~s$^{-1}$ is 
too low compared to that of runaway stars, $\sim 130$ ~km~s$^{-1}$ (Bromely et al. 2009). 
To obtain such a low velocity dispersion, the ejection mechanism would have to be directionally coherent, which is
highly unlikely. For the remaining young stars, it is difficult to distinguish, on an individual basis,
 between a runaway star and an LA member. 
This is because the RV is compatible with both the kinematics of a runaway star and the orbital motion of a
binary star in the LA. Massive young stars are known to form predominantly in binary and
 multiple systems (Sana et al. 2012), thus it is likely that our single-epoch RVs are affected.
Given these two stochastic effects on RVs --- contamination by runaway stars and binary orbital motion ---
it is remarkable that we have found five young stars exhibiting a dispersion of only 30~km~s$^{-1}$, and a 
mean RV compatible with LA kinematics.
Considering also the areal incompleteness (Casetti-Dinescu  et al. 2012),
what we have discovered  is probably the ``tip of the iceberg''.


Absolute magnitudes and ages have been derived for all the young stars based on isochrones in the 
$log~g-T_{eff}$ plane (Bressan et al. 2012).
In Figure 3, we plot distance modulus versus age for these stars. Stars with 
RV $ > 150$~km~s$^{-1}$ are once again highlighted.
As a guide, the gray band indicates the kinematical distance (21 kpc) with uncertainty, 
derived for a high velocity cloud in the LA by McClure et al (2008) that crosses
the Galactic disk. Clearly, this kinematic distance is 
within the range of our stellar distances. A number of more nearby stars are 
also present in this young sample, and they may be more readily explained as having originated in the disk.

Finally, we discuss the notable O6V star (Tab. 1). Its origin in the MW disk is doubtful, since it is 
too young (1-2 Myr) to have traveled at a reasonable speed to a Galactocentric distance of $\sim 39$ kpc. 
To demonstrate this, we calculate its orbit using a three-component Galactic potential model and find its 
last disk crossing occured over 500 Myr ago.  Even allowing for uncertainties in the distance,
the shortest time since last disk crossing in the orbit is 385 Myr.
Therefore, this star could not have been born in the Galactic disk.
Another possiblity is that it was born in the LMC and subsequently ejected, but this would
require a velocity of the order of $10^4$ ~km~s$^{-1}$ for such a young star to reach its 
current position.
This being an unrealistic value, the only viable 
possibility is that it was born {\it in situ}, far away from both the Galactic disk and the LMC.

These observations establish that conditions were met for recent star formation  
in the LA material located in the outskirts of the Galactic disk ($R\sim 18$ kpc), most likely as a 
consequence of the interaction between the Galactic disk and 
portions of the LA. We note that the most distant HI structure associated with the MW disk, is a
spiral arm at $R = 18$ to 24 kpc (McClure-Griffiths et al. 2004), while stellar samples indicate 
shorter distances for the ``edge'' of the disk of $\sim 14$ kpc (e.g., Minniti et al. 2011)

Our findings cast  new light on the interaction of the Clouds with the MW, perhaps making a 
first infall scenario less likely. Whether this is the case remains to be established by more complex models, and in light
of the lower velocity of the Clouds as indicated by the Vieira et al. (2010), Costa et al. (2009) and the Kallivayalil et al. (2013) 
studies compared to the first HST study by Kallivayalil et al. (2006).

\begin{figure}
\includegraphics[scale=1.0,angle=-90]{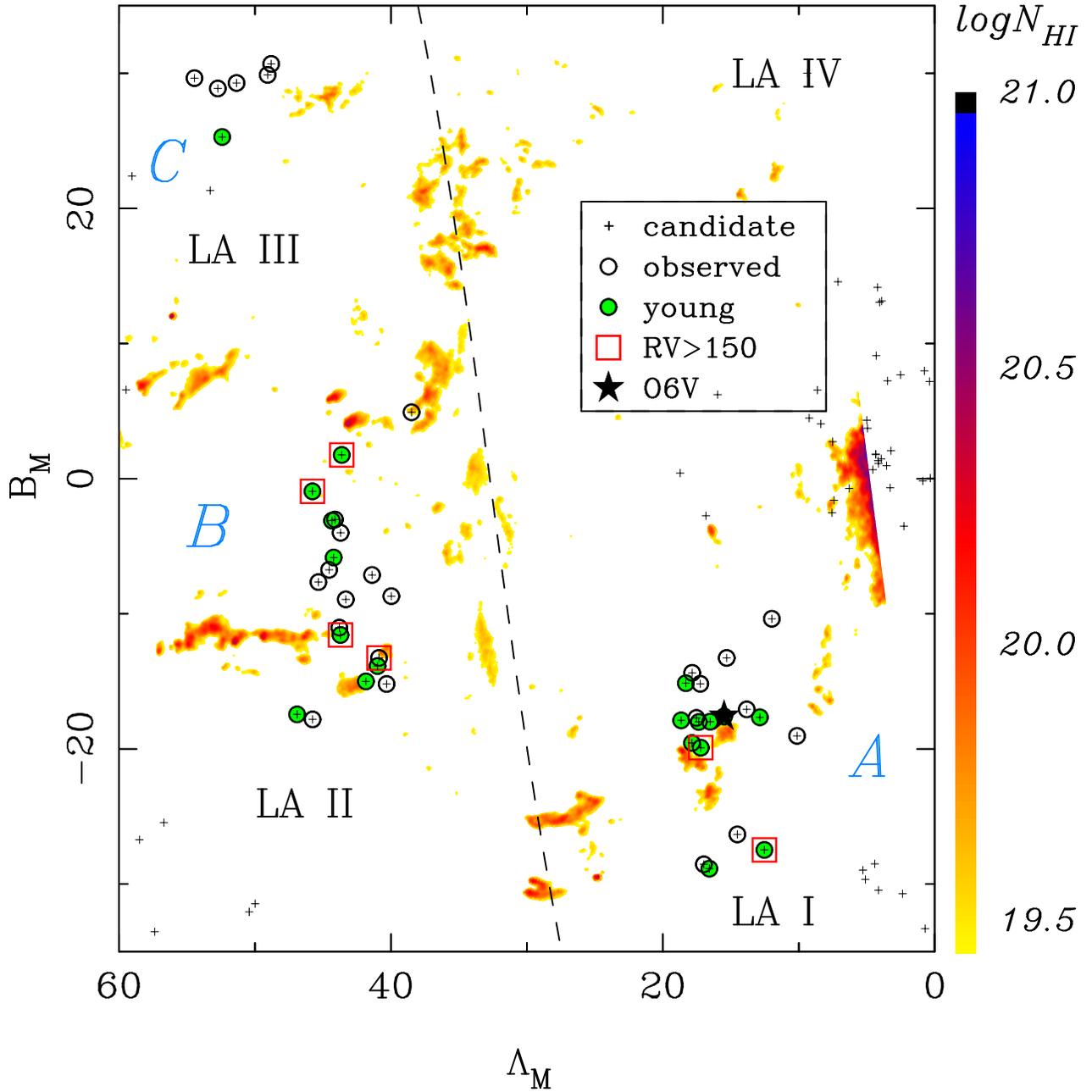}
\caption{The spatial distribution in Magellanic coordinates of our OB candidates (crosses)
The background color map shows the H I density distribution for $ 150 \le RV_{LSR} \le 400$ km/s,
with the main LA branches 
(Venzmer et al. 2012)
indicated. 
The 42 OB candidates observed spectroscopically are shown with circles.
Filled green symbols indicate the young stars, while symbols highlighted with red squares 
indicate stars with radial velocity $RV > 150$ km s$^{-1}$. The black star symbol represents
the most massive, young star in our sample (sp. type O6V). Our three 
regions of interest (A, B and C) are also labeled.  
The dashed line represents the Galactic plane.}
\end{figure}

\begin{figure}
\includegraphics[scale=1.0]{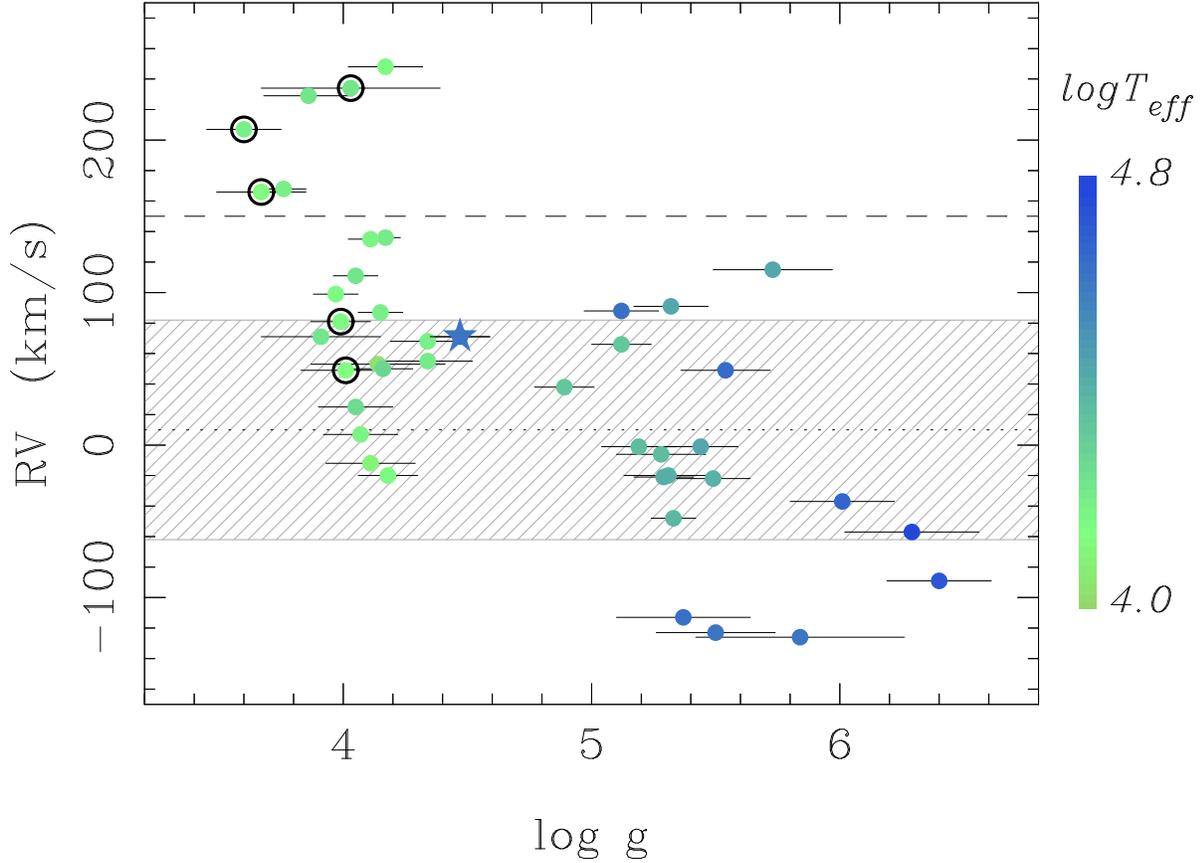}
\caption{Kinematic and spectral properties of our OB candidates.
Heliocentric radial velocity as a function of surface gravity for 42 observed stars.
Each symbols' color represents the effective temperature as indicated.
The mean and $\pm2\sigma$ standard deviation for the Galactic thin+thick disk 
(Robin et al. 2003) are indicated with a hatched area. 
The horizontal line at 150~km~s$^{-1}$ shows
the limit for LA RV-member candidates. Fast rotators (vsini $>100$~km~s$^{-1}$) are 
highlighted with a black circle. Note, the group of six stars with velocities in excess of 
150~km~s$^{-1}$ and log~g smaller than $\sim 4.2$ dex. Five of these are classified as 
massive, young stars, and only one as a sdB, primarily 
on account of its low He abundance (see Table 1). Note also the hot, relatively low
log~g star at RV $\sim 70$~km~s$^{-1}$ (star symbol). This is the earliest spectral type star found 
with this surface gravity, and thus classified as O6V, a massive, young star located at $\sim 40$ kpc
from the Sun.}
\end{figure}

\begin{figure}
\includegraphics[scale=1.0]{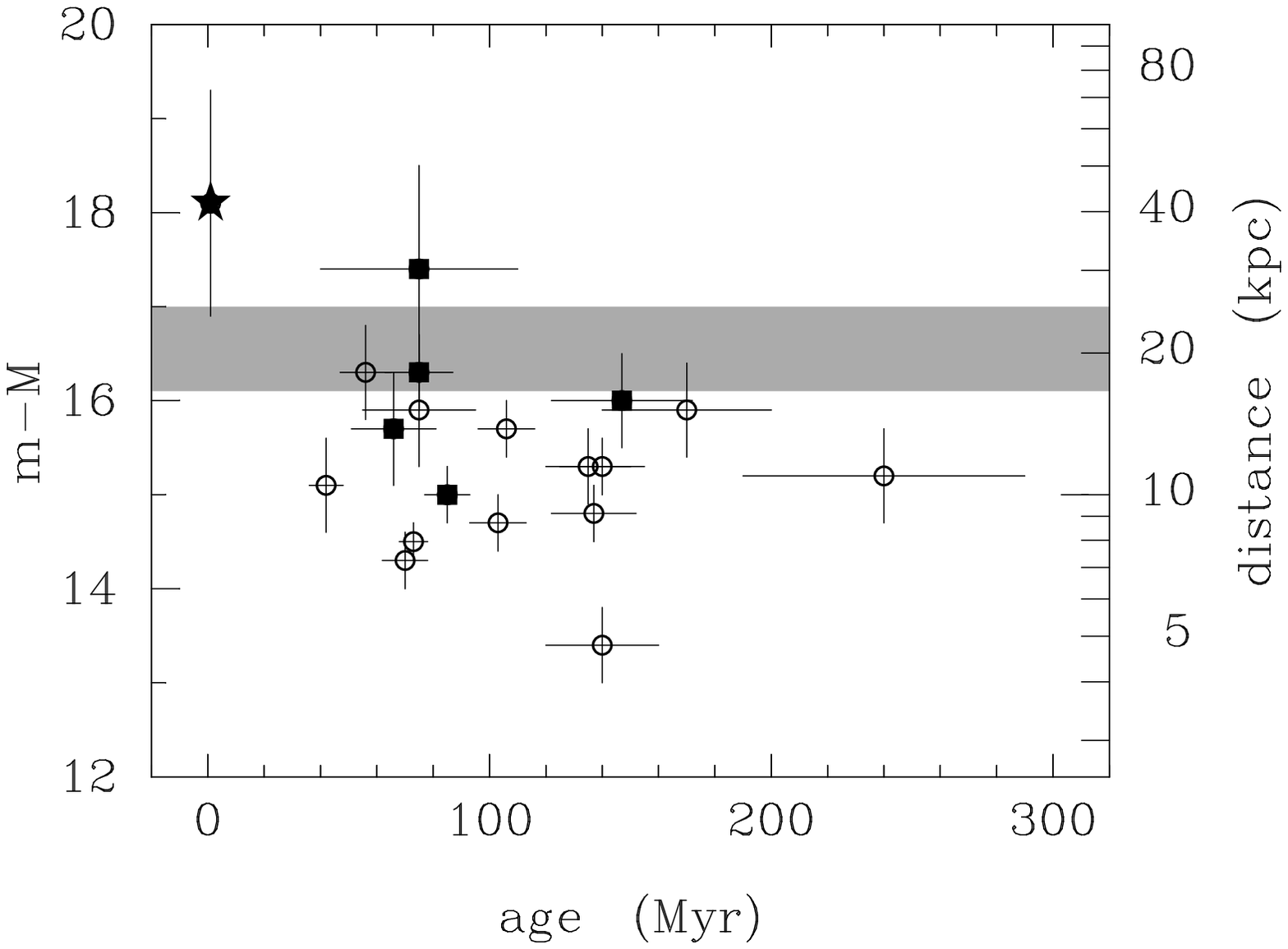}
\caption{Distances and ages for the young stars.
Distance moduli versus ages are shown for our nineteen massive, young stars.
The stars with RV$>150$ km~s$^{-1}$ are shown with filled squares. The star symbol
indicates the O6V star.
The gray band represents the kinematical distance of one high velocity cloud member of the LA
(McClure-Griffiths et al. 2008);
the width of the band corresponds to a $20\%$ error in the distance.}
\end{figure}

This investigation is based on data gathered with the 6.5-m Baade telescope, located
at Las Campanas Observatory, Chile (program ID: CN2013A-152). 
D.I.C. acknoweledges partial support by the NSF through the grant 0908996.
R.A.M. acknowledges partial support from 
Project IC120009 ``Millennium Institute of Astrophysics (MAS)'' of the 
Iniciativa Cientifica Milenio del Ministero de Economia, Fomento y Turismo de Chile, and from project PFB-06 CATA.


\begin{references}


\reference{} Baschek, B. 1975, in: Problems in stellar atmospheres and envelopes New York, Springer-Verlag New York Inc. 1975, 101-148
\reference{} Bergeron, P., Saffer, R. A., \& Liebert, J. 1992, \apj, 432, 351
\reference{} Bressan, A., Marigo, P., Girardi, L., Salasnich, B., Dal Cero, C., Rubele, S., \&
Nanni, A. 2012, \mnras, 427, 127
\reference{} Bromley, B. C., Kenyon, S. J., Brown, W. R., \& Geller, M. J. 2009, \apj, 706, 925
\reference{} Casetti-Dinescu, D. I., Vieira, K., Girard, T. M., \& van Altena, W. F. 2012, \apj, 753, 123
\reference{} Costa, E., M\'{e}ndez, R. A., Pedreros, M. H., Moyano, M., Gallart, C., Noel, N., Baume, G. \& Carraro, G. 2009,
\aj, 137, 4339.
\reference{} Demers, S., Battinelli, P.  1998, \aj, 115, 154
\reference{} Diaz, J. D. \& Bekki, K. 2011, PASA, 28, 117
\reference{} Diaz, J. D. \& Bekki, K. 2012, \apj, 750, 36
\reference{} For, B., Staveley-Smith, L. and McClure-Griffiths, N. M. 2013, \apj, 764, 74
\reference{} Geier, S. \& Heber, U. 2012, \aap, 543, 149
\reference{} Irwin, M. J., Demers, S., \& Kunkel, W. E.  1990 \aj, 99, 191
\reference{} Kalberla, P. M., W., {\it et al.} 2010, \aap, 521, 17
\reference{} Kallivayalil, N., {\it et al.} 2006, \apj, 638 772
\reference{} Kallivayalil, N, van der Marel, R. P., Besla, G., Anderson, J., \& Alcock, C. 2013, \apj, 764 161
\reference{} Maxted, P. F. L., Heber, U., Marsh, T. R., \& North, R. C. 2001, \mnras, 326, 1391
\reference{} McClure-Griffiths, N. M., Dickey, J. M., Gaensler, B. M. \& Green, A. J. 2004, \apj, 607, L127
\reference{} McClure-Griffiths, N. M., {\it et al.}, 2008, \apj, 673, L143
\reference{} McClure-Griffith, N. M., Staveley-Smith, L., Lockman, F. J., Calabretta, M. R., Ford, H. A., 
Kalberla, P. M. W., Murphy, T., Nakanishi, H., and Pisano, D. J., 2009, \apjs, 181, 398 
\reference{} Minniti, D., Saito, R. K., Alonso-Garcia, J., Lucas, P. W., \& Hempel, M. 2011, \apj, 733, L43
\reference{} Moni Bidin, C., Villanova, S., Piotto, G., \& Momany, Y. 2011, \aap, 528, 127
\reference{} Moni Bidin, C. {\it et al.} 2012 \aap, 547, 109
\reference{} Munari, U., Sordo, R., Castelli, F., \& Zwitter, T. 2005, \aap, 443, 1127
\reference{} Napiwotzki, R., Green, P. J., \& Saffer, R. A. 1999, \apj, 517, 399
\reference{} Nidever, D., Majewski, S. R., Burton, W. B., \& Nigra, L. 2010, \apj, 668, 949
\reference{} Robin, A. C., Reyl\'{e}, C., Derri\`{e}re, S., \& Picaud, S. 2003, \aap, 409, 523
\reference{} Saffer, R. A., Bergeron, P., Koester, D. \& Liebert, J.  1994, \apj, 432, 351
\reference{} Sana, H. {\it et al.} 2012, Science 337, 444
\reference{} Tonry, J. \& Davis, M. 1979, \aj, 84, 1511
\reference{} Venzmer, M. S., Kerp, J., \& Kalberla, P. M. W. 2012, \aap, 547, 12
\reference{} Vieira, K. {\it et al.} 2010, \aj, 140, 1934




\end{references}
\end{document}